\documentclass[doublecol]{epl2}
\usepackage{graphicx}
\usepackage{amsmath}

\begin{document}

\title{A NOT operation on Majorana qubits with mobilizable solitons in an extended
Su-Schrieffer-Heeger model}

\author{Ye Xiong \inst{1} \and Peiqing Tong \inst{1,2,3}}

\institute{
  \inst{1} Department of Physics and Institute of Theoretical Physics
  , Nanjing Normal University, Nanjing 210023,
P. R. China \\
  \inst{2} Jiangsu Key Laboratory for Numerical Simulation of Large
  Scale Complex Systems, Nanjing Normal University, Nanjing 210023,
P. R. China \\
  \inst{3} Kavli Institute for Theoretical Physics China, CAS,
Beijing 100190, China
}

\pacs{73.20.-r}{Electron states at surfaces and interfaces}
\pacs{73.43.-f}{Quantum Hall effects}
\pacs{03.65.Vf}{Phases: geometric; dynamic or topological}
\pacs{71.10.Pm}{Fermions in reduced dimensions}

\abstract{
  Coupling Majorana qubits with other qubits is an absolute essential in
  storing, manipulating and transferring informations for topological
  quantum computing. We theoretically propose a manner to coupling
  Majorana qubits with solitons, another kind of topological impurities,
  which was first studied in the spinless Su-Schrieffer-Heeger (SSH)
  model. We presents a NOT operation on the Majorana qubit
  with moving the complementary soliton through heterostructure
  adiabatically. Based on these two topological impurities, the
  operation is robust against local disorder.  Furthermore, we find that
  the soliton may carry decimal electric charge instead of fractional
  charge $1/2$, because of the breaking of gauge invariance induced by
  superconducting proximity.
}

\maketitle

\section{Introduction}

Topological phases (TP) of matter are characterized by nontrivial band
structures which can not be connected to trivial band structures without
closing the band gap at the Fermi energy \cite{RevModPhys.83.1057,
RevModPhys.82.3045, RevModPhys.80.1083, TIbook}. Due to the holographic
principle, symmetry protected boundary states will appear at the
boundaries of the system. When the energy gap is opened by
superconductivity, the superconductor with nontrivial band structure
becomes topological superconductor and Majorana fermions (MFs) may exist
at the surfaces.  The $Z_2$ invariant which corresponds to the parity of
the number of the MF branches at each boundary can be used to
distinguish the superconductive nontrivial TP from the superconductive
trivial TP \cite{RevModPhys.82.3045, PhysRevLett.95.146802,
PhysRevB.75.121306}.  These MFs are robust against local distortions and
are considered to be suitable for physical realization of topological
qubit \cite{RevModPhys.80.1083}.  There are many proposals to generate
topological superconductors hosting MFs, based from 1-dimensional (1D)
superconducting wires with strong spin-orbital interaction \cite{Kitaev,
PhysRevLett.105.077001} to 2D topological superconductor
heterostructures \cite{PhysRevLett.100.096407, PhysRevB.81.125318} or
vortex cores \cite{PhysRevB.79.094504}.

To realize quantum computing based on Majorana qubits, one needs to
transfer quantum informations between different qubits. This is a
challenging work because the nonlocal nature of MFs prohibits local
operator to coherently transfer quantum information into and out of
topological systems. There are already some proposals to hybridize
Majorana qubits with other qubits, for instance, with fluxonium
qubit\cite{PhysRevLett.111.107007}, with flux
qubit\cite{PhysRevLett.106.130504}, with quantum dot
qubit\cite{PhysRevLett.106.130505} and with superconducting charge
qubits\cite{been1}. In this paper, we propose to couple the Majorana
qubits with soliton qubit, another kind of topological impurities first
studied in the spinless Su-Schrieffer-Heeger (SSH) model.  These soliton
qubits are totally different from the qubits enrolled in the previous
proposals. They are topological, localized and mobilizable. In this
paper, we find that moving the complementary soliton adiabatically
through the SSH region of a heterostructure can induce a NOT operation
on the Majorana qubit. This manipulation will operate one kind of
topological qubit with another kind of topological qubit. So the
operation should be fault tolerant and robust against local disorder. 

We start from a 1D extended spinless Su-Schrieffer-Heeger (SSH) model.
An effective nearest neighboring $p$ wave superconducting pairing is
involved so that the model can be considered as a combination of the SSH
model \cite{PhysRevLett.42.1698,
TIbook} and the Kitaev's toy model \cite{Kitaev}. As varying the
parameters, the phases of the model can evolute from TP hosting MFs
to another TP hosting solitons. Because the present of superconducting
pairing, the electric charge carried by each soliton is no longer
universally equal to $1/2$ (in the units $e=1$). We will raise a
topological way to calculate this charge accurately in Sec. 3.

We begin in Sec. 2 by introducing a 1D tight-binding
model. This model has three topological nonequivalent phases. We will
also illustrate the kinds of topological impurities in these phases.
In Sec. 3, we enroll the Thouless pump to the model and
calculate the movement of Wannier functions (WFs) during the pump. This
can help us find the electric charge carried by each soliton.
In Sec. 4, we propose how to apply NOT operation on a
Majorana qubit by moving the complementary soliton adiabatically and
discuss why local disorder can not affect the operation. 
In Sec. 5, we emphasis the importance of our findings and
discuss new vistas of research in this field.

\section{1D tight-binding model and phase diagram}\label{section2}

\subsection{Hamiltonian}
We use a Hamiltonian describing 1D spinless dimerized model with
the nearest neighboring $p$ wave superconducting coupling,
\begin{eqnarray}
  H &=&\sum_i \{[1+(-1)^i\delta] c^\dagger_i c_{i+1} +\text{h.c.} \}+h \sum_i
c^\dagger_i c_i   \nonumber \\ 
& & +\Delta \sum_i (c^\dagger_i c^\dagger_{i+1}
+\text{h.c.}),
  \label{Hamiltonian}
\end{eqnarray}
where $c^{\dagger}_i$ and $c_{i}$ are the creation and annihilation
operators for electron on site $i$, respectively, $\delta$ is the degree
of dimerization, $\Delta$ is the strength of $p$ wave pairing, and $h$
is the external global potential. We will only investigate the model with
$|\delta|<1$.

This model could be realized by placing a dimerized polyacetylene on top
of an $s$-wave superconductor. A rotating magnetic field can induce the
required giant spin-orbital interaction along the chain
\cite{PhysRevX.3.011008}. The effective spinless Hamiltonian Eq.
(\ref{Hamiltonian}) is obtained when the Fermi energy lying in the gap
opened by the staggered hoppings or by the external magnetic field.
Another possible realization of the Hamiltonian has been proposed by
Klinovaja {\it et al.} \cite{PhysRevLett.109.236801}, in which a helical
magnetic field plays the role of staggered hoppings.

\subsection{Phase diagram}

\begin{figure}[ht]
  \includegraphics[width=0.45\textwidth]{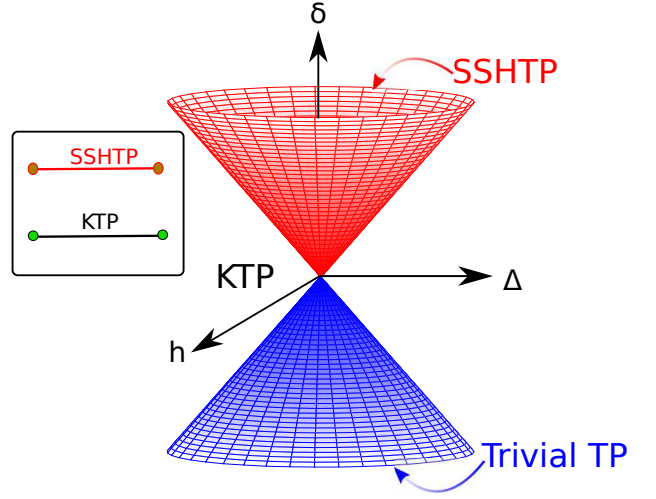}
  \caption[Fig]{\label{fig1} The phase diagram in the parameter space
  spanned by $\delta$, $\Delta$ and $h$. The system is in the SSHTP in 
  the region enclosed by the red cone. Within the blue cone and in
  the region outside two planes $h=\pm2$ (not showed in the figure), it
  is in the trivial TP. The system is in the KTP in the rest parameter
  region. The left inset shows that an open chain in the SSHTP (or in
  the KTP) can have soliton states (or MFs) as the topological impurities at the ends.}
\end{figure}

The phase boundaries separating different phases in the phase diagram
are determined by the fact that the band gap closes there.
The Hamiltonian in momentum $k$ reads,
\begin{eqnarray}
  H(k) &=
  & \{[(1-\delta)+(1+\delta)\cos(k)]\sigma_x+(1+\delta)\sin(k)\sigma_y\
  \nonumber \\
  & +&h\sigma_0 \}\otimes\tau_z-\Delta\tau_y\otimes[\sin(k)\sigma_x +(1-\cos(k))\sigma_y] 
  \nonumber
  \label{Hamilk}
\end{eqnarray}
The Pauli matrix $\tau_{x,y,z}$ and
$\sigma_{x,y,z}$ are operating on particle-hole and 
sub-lattice subspaces, respectively. $\sigma_0$ is a unit matrix.The band gap closes when the
parameters obey one of the following two conditions, $\frac{h^2}{4}
+\Delta^2 =\delta^2$(when $|\delta|<1$) and $h=\pm2$. In Fig.
\ref{fig1}, we sketch out the first condition with the
ellipsoidal cones in the parameter space spanned by $h$, $\delta$ and
$\Delta$. Two planes at $h=\pm2$ (for the second condition) are not
showed for the sake of clarity.

From the phase diagrams for the SSH model and Kitaev's toy model, one
can anticipate the topological properties of the TPs in our model. In
Fig. \ref{fig1}, for the parameter region enclosed by the red cone, the
system is in the TP similar to that of the SSH model. This is because
the model can be
regressed to the standard SSH model by decreasing both $h$ and
$\Delta$ to $0$ while keeping $\delta>0$. The band gap at
the Fermi energy does not close during this regression. So we call the
TP in the red cone as a SSH like TP (SSHTP).  For the parameter region
in between two ellipsoidal cones and two planes $h=\pm 2$ at distance,
the system is in the Kitaev like TP (KTP) because it can be regressed to
a standard Kitaev's toy model in TP without closing the band gap. In the
left inset of Fig.  \ref{fig1}, we schematically show that an open chain
in  SSHTP (or KTP) can host soliton states (or MFs) at the ends. For the
rest regions in the parameter space, including that enclosed by the blue
cone and those outside two planes $h=\pm2$, the system is in the trivial
TP.  Actually, the trivial phase in these regions can be separately
regressed to the trivial TP of the SSH model and to the trivial TP of
the Kitaev's toy model, respectively.

\subsection{Topological impurities appeared at boundaries}

\begin{figure}[ht]
  \includegraphics[width=0.45\textwidth]{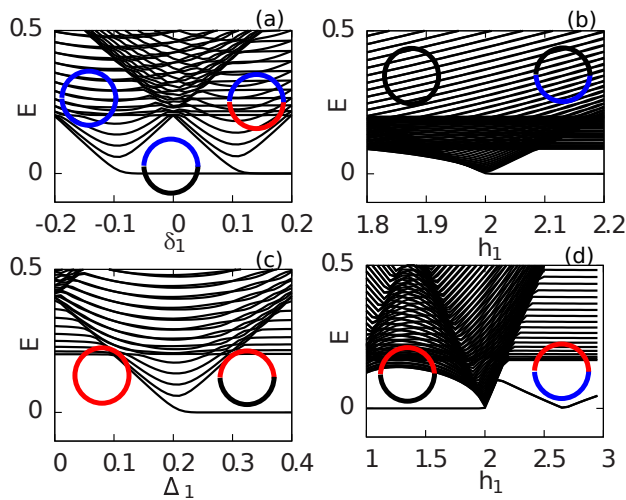}
  \caption[Fig]{\label{fig2} The energy spectrum of a ring in which the
  upper half and lower half are in different parameter regions. The
  length is $N=400$. The
  parameters are: (a) $h=0$, $\Delta=0.1$ and $\delta=-0.2$ in the upper
  half and $h=0$, $\Delta=0.1$ and $\delta=\delta_1$ in the lower half;
  (b) $h=1.8$, $\Delta=0.1$ and $\delta=0.2$ in the upper half and
  $h=h_1$, $\Delta=0.1$ and $\delta=0.2$ in the lower half; (c) $h=0$,
  $\Delta=0.1$ and $\delta=0.2$ in the upper half and $h=0$,
  $\Delta=\Delta_1$ and $\delta=0.2$ in the lower half; (d) $h=0.6$,
  $\Delta=0.1$ and $\delta=0.4$ in the upper half and $h=h_1$,
  $\Delta=0.1$ and $\delta=0.4$ in the lower half.
  The phases of the upper and lower parts of the ring are
  indicated by colors: red for SSHTP, black for KTP and blue for trivial TP.}
\end{figure} 

To further illustrate the topological nonequivalence of the $3$ phases
in the phase diagram and show the topological impurities appeared at
boundaries, in Fig.  \ref{fig2}, we plot the energy spectrum of a ring,
in which parameters are adjusted so that the upper and lower parts of
the ring are in different phases. The topologically protected boundary
states at the joints of two parts should emerge in the energy spectrum
with energies inside the band gap.  In all panels of Fig. \ref{fig2},
the parameters are fixed in the upper half of the ring, while one of the
parameters,referred to $\delta_1$ in (a), $h_1$ in (b) and (d), and
$\Delta_1$ in (c), is varying in the lower half.  We indicate types of
TPs of the upper and lower parts with different colors: red, black, and
blue that represent SSHTP, KTP and trivial TP, respectively. In Fig.
\ref{fig2}(a), by varying staggered hoppings $\delta_1$ in the lower
half, the ring changes from full trivial TP to half trivial TP and half
KTP at $\delta_1=-0.1$. As a result one zero energy state representing
the emergency of two MFs at the joints appears after this transition.
When $\delta_1$ is further increased and exceeds $+0.1$, the lower part
of the ring enters SSHTP.  So in this case, there are two zero energy
states inside the band gap that representing the existence of two
soliton states. In Fig.  \ref{fig2}(b), when $h_1<2$ the parts are in
the KTP and there is no zero energy state in the gap because there is no
boundary between different phases.  But when $h_1>2$, the phase of the
lower half changes to the trivial TP, and we find a zero energy state in
the spectrum. These results can be traced back to the well known
conclusions in the standard SSH model or in the Kitaev model. Something
new happens in the next panels.  In Fig.  \ref{fig2}(c), the ring is
half in the SSHTP and half in the KTP when $\Delta_1>0.2$. At each
joint, only one MF is left because the coupling between MF and soliton
state at the joint lifts the energy of the soliton state. So there is
only one zero energy state left in the gap.  In Fig.  \ref{fig2}(d), the
ring is half in the SSHTP and half in the trivial TP when $h_1>2$. This
situation is same as that in the right part of Fig.  \ref{fig2}(a). But
the soliton states in (d) are lifted from zero energy and become Andreev
bound states. This Andreev bound states are different from the normal
Andreev bound states because they take the responsibility of topological
impurities which must appear at the boundaries of different TPs. As a
result, as the panel (d) shows, the Andreev bound states can evolves
smoothly to zero energy states without closing the band gap.

\begin{figure}[ht]
  \includegraphics[width=0.4\textwidth]{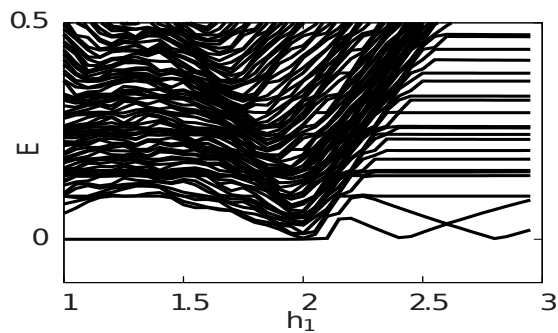}
  \caption[Fig]{\label{fig3} The energy spectrum of a disordered
  ring. Most of the parameters are the same as those in Fig. \ref{fig2}(d)
  except the randomized staggered hoppings $t_i=1+ (-1)^i
  \delta +0.15w$, where $w$ is a random number in $[-1,1]$.}
\end{figure}

Because of this topological nature, the evolutions to zero energy
states should be robust against disorder. In Fig.  \ref{fig3}, disorder
is enrolled to the ring by replacing the staggered hopping between the
$i$th and the $(i+1)$th site with $1+ (-1)^i \delta + 0.15w$, where $w$
is random numbers distributing uniformly in $[-1,1]$.  Other parameters
are the same as those in Fig. \ref{fig2}(d). The figure shows that disorder
can remove degeneracies of the Andreev bound states but can {\it not}
destroy them.

\section{Thouless pump and decimal charge carried by each
soliton}\label{section3}

Similar to that in the standard SSH model, domain walls are mobilizable
in the model. To electrically control its movement, the electric charge
carried by each domain wall should be determined at the first. In this section, with the help of
the evolution of Wannier functions(WFs) during the Thouless pump, we can
determine the charge accurately.

\subsection{Wannier functions for the occupied bands}
The most localized WFs for the occupied bands are defined as the 
eigenvectors of the tilde position operator
\begin{equation}
  \tilde R = \hat P \hat R \hat P,
  \label{Roperator}
\end{equation}
where $\hat R$ is the position operator and $\hat P$ is the projection
on the occupied states \cite{ThoulessWannier, PhysRevB.74.235111,
PhysRevLett.107.126803}. Here $\hat P$ can be written explicitly as
$\hat P = \sum_{\alpha \in \text{occupied states}} |\alpha\rangle
\langle \alpha|$ and
in the site representation the position operator $\hat R
=\text{diag}(1,2,\cdots,N)\tau_0$, where $\tau_0$ is the $2\times2$ unit
matrix in the particle-hole subspace and $\text{diag}(1,2,\cdots,N)$ is
a diagonal matrix with the diagonal elements running through lattice
sites from $1$ to $N$. The eigenvalues of $\tilde R$, denoted as $R$s,
are the central positions of the WFs. Eq. \ref{Roperator} extends
the projected position operator to particle and hole
subspaces. As a result, the WFs for the unoccupied bands (from holes),
which are absent in the usual particle representation, can be obtained from this
new definition. The usual WFs obtained from the traditional definition
are plotted in panel (c) in Fig. \ref{fig4} for comparison.

\subsection{Thouless pump}
The Thouless pump is
introduced to the Hamiltonian by varying the staggered hopping and
on-site energy slowly with an extra parameter $\phi$ \cite{TIbook}. Then
the Hamiltonian becomes
$
  H(\phi)=\sum_i [1+(-1)^i\delta\cos(\phi)] c^\dagger_i c_{i+1}
  +\text{h.c.}+\sum_i (-1)^i h_\text{st}\sin(\phi) c^\dagger_i c_i
  + h\sum_i c^\dagger_i c_i
  +  \Delta \sum_i (c^\dagger_i c^\dagger_{i+1}
+\text{h.c.}).
$

\subsection{Evolution of WFs during the Thouless pump}

\begin{figure}[ht]
  \includegraphics[width=0.45\textwidth]{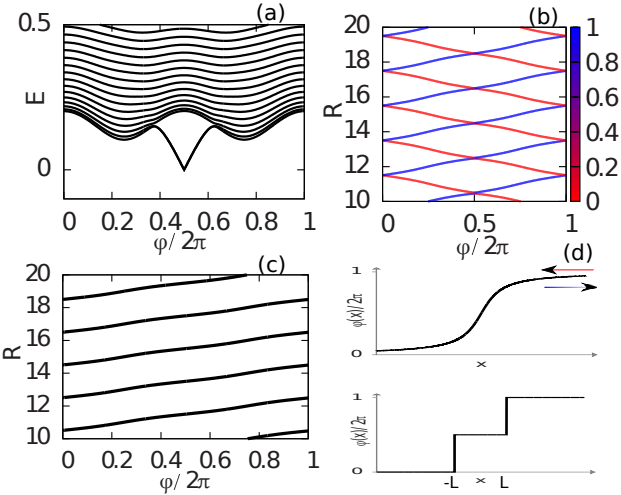}
  \caption[Fig]{\label{fig4} The energy spectrum (a) and the WFs centers
  (b) during the Thouless pump of $\phi$. The parameters are $h=0$,
  $h_{\text st}=0.3$, $\delta=-0.2$ and $\Delta=0.1$. The colors of the
  points in panel (b) represent the particle weights of the
  corresponding WFs in the particle-hole subspace with scale given by
  the right palette, where $1$ (blue) means that the WF contains one
  particle while 0 (red) indicates that the WF is completely hole like.
  (c) For comparison, the conventional WFs for a standard SSH model
  during Thouless pump are showed. (d) A schematic shows the two kinds of layouts of $\phi(x)$
  along chains for the arguments in the next subsection.} 
\end{figure} 

In Fig. \ref{fig4}, we plot the energy spectrum (a) and the centers of
WFs (b) as functions of $\phi$ for a $N=200$ chain with open boundary
condition. The parameters are $h=0$, $h_\text{st}=0.3$, $\delta=-0.2$
and $\Delta=0.1$. As Fig. \ref{fig4}(a) shows, the chain is in trivial
TP with no boundary state around $\phi=0$.  As $\phi$ is varying, the
chain is pushed to the SSHTP near $\phi=\pi$ and then is pulled back to
the trivial TP at the end of a cycle. During this pump, the band gap at
the Fermi energy keeps open so that the WFs we calculated are localized
and their centers showed in (b) are reliable. As the model is
represented in the Nambu representation, an extra new set of WFs
corresponding to the bands of holes emerges.  In Fig. \ref{fig4}(b), we
use the colors, blue and red to illustrate the weights of the WFs on the
particle and hole subspaces, respectively.  So blue(red) points in the
panel are illustrating that WFs centered at the positions are particle(hole) like
functions. Panel (b) shows that the particle like WFs are moving right
accompanied with the hole like WFs moving left during the pump.

\subsection{Electric charge carried by each soliton}

For the standard SSH model, there is a counting formula describing the total
number $M$ of the unoccupied zero energy states in term of electric
charge $Q$ carried by each topological impurity (including both domain
wall and geometric end), $M=-2Q \text{mod} 2$ \cite{PhysRevLett.46.738,
Jackiw1981253}.  The well known fractional charge carried by each
soliton is a direct conclusion of this equation (when $M=1$, $Q=
\frac{1}{2}$).  Turning on the superconducting coupling will not break
the charge conjugation symmetry and the counting formula should survive.
But the electric charge $Q$ must be replaced by the conserved
quasiparticle charge $Q_{\text{BdG}}$ \cite{PhysRevB.83.104522} because
the global electromagnetic gauge invariance is broken in the BdG mean
field Hamiltonian. It should be emphasized that $Q_{\text{BdG}}$ is a
topological character in the present case and has nothing to do with the
actual electric charge $\tilde{Q}$ anymore.  So we need to figure out
the actual charge $\tilde{Q}$ carried by each soliton in order to
electrically control their motion.

We use a thought experiment to figure out the actual electric charge
$\tilde{Q}$ as well as the conserved BdG charge $Q_{\text{BdG}}$ carried
by each soliton.  Suppose that there is an infinite chain with
Hamiltonian of Eq. (\ref{H_cp}).  The parameter $\phi$ is not fixed but
varying slowly along the chain as $\phi(x)$. The total variation of
$\phi$ along the chain is $2\pi$. Without loss of generality, we let
$\phi(-\infty)=0$ and $\phi(+\infty)=2\pi$.  As $\phi(x)$ is varying
slowly (showed in the upper panel in Fig. \ref{fig4}(d)), the positions
of  WFs for $\phi(x)$ given by Fig.  \ref{fig4}(b) are still
valid in regions in which $\phi(x)$ changes slightly. Now let's
compare the positions of WFs in this chain with those in a uniform chain
in which $\phi(x)$ is fixed at $0$. At the far left segments of the
chains, the positions of WFs in the two chains are the same because
$\phi(x)=0$ in both regions. As increasing $x$, the WFs are moved
slightly away from the positions for the uniform chain because $\phi(x)$
increases. As Fig. \ref{fig4}(b) shows, particle like WFs are misaligned
in the $x$ direction accompanied with hole like WFs misaligned in the
inverse direction. These deviations keep increasing with $x$ and reach
the length of one unit cell in the far right region in which
$\phi(x)=2\pi$.  So we conclude that, compared with the uniform chain,
the chain with varying $\phi(x)$ loses a particle like WF and gains a
hole like WF. This can also be looked as that $\phi(x)$ pushes out a
particle like WF and pulls in a hole like WF at the right end of the
chain at $x=\infty$.  Now we relax the restriction that $\phi(x)$ is
varying slowly with $x$.  This relaxation does not alter the above
conclusion because the local variation at finite $x$ will not affect
physics at $\infty$.  We choose the new layout of $\phi(x)$ as
$\phi(x)=0$ when $x<-L$, $\phi(x)=\pi$ when $L>x>-L$ and $\phi(x)=2\pi$
when $x>L$, where $L$ is a large but finite number (showed in the lower
panel in Fig. \ref{fig4}(d)). The new layout is representing a chain
with a pair of domain wall and anti-domain wall at $\pm L$. We know that
each domain wall (anti-domain wall) has a soliton state on it. So these
two soliton states must take the responsibility of the lost and gained
WFs. As a result, each soliton takes half of the total lost charges,
$\tilde{Q}=-\frac{1}{2}[\langle \Psi_{\text{Particle like WF}}|\hat \rho
| \Psi_{\text{Particle like WF}}\rangle - \langle \Psi_{\text{Hole
like WF}}|\hat \rho |\Psi_{\text{Hole like WF}}\rangle]$, where
$\hat\rho$ is the single particle
density operator $\hat\rho = \sum_i c^\dagger_i c_i$ and
$|\Psi_{\text{Particle(Hole) like WF}}\rangle$ is a wave function of
the particle(hole) like WF. For a chain with
the parameters showed in Fig.  \ref{fig4}(a) and (b), this charge
$\tilde{Q}$ is still $\frac{1}{2}$. As to the
conserved BdG charge $Q_\text{BdG}$, this conserved charge counts that
how many WFs have been pushed out from the chain with one domain wall.
From the above argument, we see it is universal $\frac{1}{2}$.

We have calculated the energy spectrum as well as the motions
of WFs during the Thouless pump for a chain with nonzero external
potential, $h=0.3$, for comparison. Other parameters are same as those
in Fig. \ref{fig4}. We find that $h$ lifts the energies of the soliton
states at $\phi=\pi$ but the motions of WFs are
same as those showed with $h=0$. But these WFs have been
altered from pure particle like or hole like functions to mixed
functions. We can count the total charge being pump out and find that it
is $0.97$ instead of $1$. As a result, each soliton in this model carries 
decimal charge $0.485$. 
So we can conclude that the charge carried by each soliton is
{\bf not} a universal fractional number but depends on $h$ when $h\ne0$.

\section{Manipulating Majorana qubit by moving soliton}\label{section4}

\begin{figure}[ht]
  \includegraphics[width=0.45\textwidth]{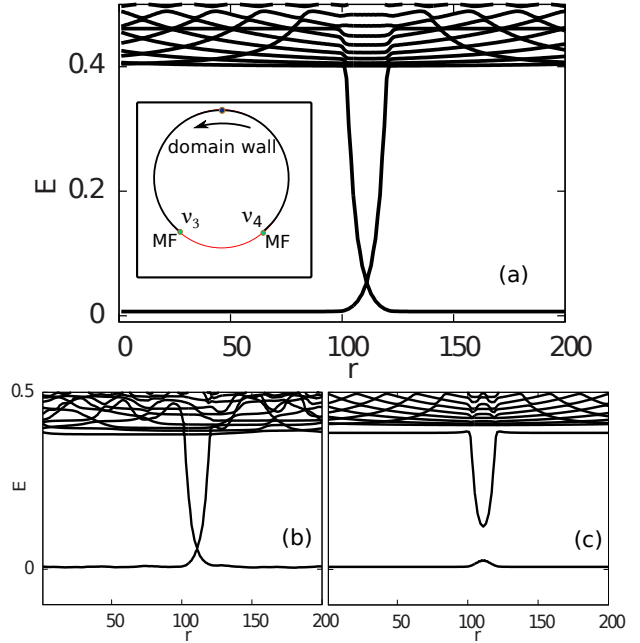}
  \caption[Fig]{\label{fig5} (a) The energy spectrum of the heterostructure
  ring, showed in the inset, when the domain wall (blue point) is
  moving.
  The length of the ring which hosts the domain wall is $N=201$. One part of
  the ring is in the KTP (black) and the rest is in the SSH phase
  (red). MFs appearing at the joints are indicated by the green points.
  The parameters are $h=0$, $\delta=0.2$, $\Delta=0$ for the part
  in the SSH phase and $h=0$, $\delta=0.2$, $\Delta=0.4$ for the other
  part. The length of the SSH region is $20$. (b) The robust level cross
  for a disordered chain. The parameters are same as those in (a) 
  except the randomized staggered hopping. (c) The level crossing
  becomes anti-crossed when $h=0.1$.}
\end{figure} 

In the present model, there are two kinds of topological impurities,
solitons and MFs. We will show that when coupling these two kinds of
topological impurities together, we can change the state of Majorana
qubit by adiabatically moving the complementary soliton along the chain.
In the following, we are concentrating on the two MFs taking part in the
coupling and ignoring the two uncoupling MFs at the other far ends.

\subsection{The heterostructure}

As the inset in Fig. \ref{fig5}(a) shows, a circular polyacetylene is
placed on the surface of a nonuniform superconductor so that one part of
the circle is in the KTP (the black arc) and the rest is in the SSH
phase (the red arc). We suppose that there is a domain wall (sketched by
the blue point) in the circle. In our numerical calculation, this domain
wall is simulated by two adjacent stronger bonds with hoppings $t+\delta$.
It will induce a soliton state when the domain wall is in the SSH
region. It should be noted that because the existence of domain wall the
total length of circle is an odd number instead of even ones.  There are
two MFs (indicated by the green points) at the joints of the arcs.

Fig. \ref{fig5}(a) shows the energy levels of the ring as the domain wall
is moved adiabatically. The domain wall starts from 
where it is showed in the inset. When it is in
the black arc in the KTP, there is only one zero energy state referring
to the two MFs because domain wall in the KTP region can not induce
soliton state inside the band gap \cite{explain_wallinktp}.  Soliton
state will appear in the gap and its energy is decreasing when the
domain wall is moved into the SSH region, while the zero
energy state is lift up in energy. These two energy levels cross when
the domain wall is passing the center of the SSH region. Because of this
level crossing, particle occupations on the zero energy state and on
the soliton state will exchange after passing the wall through the SSH region
adiabatically. This mechanism can modify the Majorana
zero energy state with a complementary soliton, say $n^{ZE}_f=n^{s}_i$ and
$n^{s}_f=n^{ZE}_i$, where $n^{ZE(s)}_{i(f)}$ is number of particle on
the MF zero energy (soliton) state before(after) the movement of domain wall.

\begin{figure}[ht]
\includegraphics[width=0.45\textwidth]{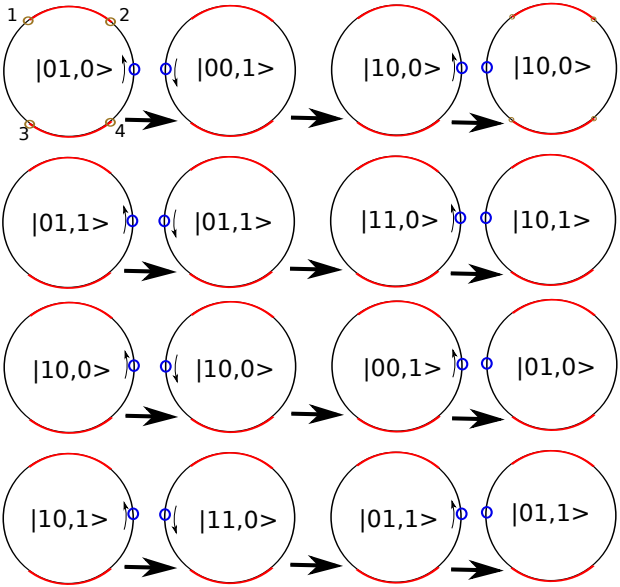}
\caption[Fig]{\label{fig6} How NOT operation works for a MF qubit. The
red region is in the SSH phase and the black region is in the KTP. The
domain-wall is the blue point. The quantum state in each case is
written at the center of each circle.}
\end{figure}

The above level crossing implies a mechanism to manipulating MF qubits
by moving soliton. A MF qubit need four MFs. Let's denote their MF operators as $\mu_1$, $\mu_2$, $\mu_3$
and $\mu_4$. The heterostructure is extended from that has been
discussed in Fig. \ref{fig6} with the SSH region (red) and the KTP region
(black). The four MFs are illustrated by $1$, $2$, $3$ and $4$ in the figure.
Our discussion is in fermion representation with the complex fermion
creation operator $\psi^\dagger_\beta=(\mu_1 +i \mu_2)/2$ and
$\psi^\dagger_\alpha= (\mu_3 +i \mu_4)/2$. The quantum states of the MF
qubit are illustrated as $|n_\alpha n_\beta \rangle$, where $n_\alpha$
and $n_\beta$ can take $0$ (empty) and $1$ (occupied). Traditionally,
the two states of a MF qubit are defined as "0" state: $|10\rangle$
and "1" state: $|01\rangle$. 

We can extend the above notation of quantum states to include the occupation on
the soliton state as $| n_\alpha n_\beta, n_{\text soliton} \rangle$.
The states showed in the figure are in this extended form. Now we show
explicitly how NOT operation works.  Supposing the initial state is
$|01,0\rangle$ with the domain wall at the "3" o'clock direction. After
moving the it to the "9" o'clock direction, $n_\beta$ and $n_{\text
soliton}$ exchange due to the level crossing.  So we get the state
$|00,1\rangle$. After moving the domain-wall a circle back to the "3"
o'clock direction. $n_\alpha$ and $n_{\text soliton}$ exchange and we get
$|10,0\rangle$. Finally the domain-wall is moved an extra one half
circle to the "9" o'clock direction. The state keeps at $|10,0\rangle$.
This is a NOT operation that changes from "1" state to "0" state for MF
qubit. In the next three rows in the figure, we show explicitly that
this NOT operation is independent on the initial state of $n_{\text
soliton}$ and works well for the case, "0" to "1". 

The only mechanism that will introduce fault to the NOT operation is
that the quasi-particle on soliton state may spontaneously jump to the
zero energy state when the zero energy state is empty. But thanks to the
localization nature of zero energy state (localized at two positions of
MFs) and soliton state, this possibility is in the scale of $1/L$ in
most case, where $L$ is the length of the KTP region.  

\subsection{Robust level crossing against disorder}

This NOT operation on Majorana qubit depends crucially on the fact that
the level crossing must be robust against disorder or must be protected.
In Fig. \ref{fig5}(b), we plot the energy spectrum for a disordered ring.
The parameters are the same as those in (a) except that the
staggered hoppings are randomized, $t_i=1+(-1)^i \delta +0.1w$, where
$w$ is a random number in $[-1,1]$.  This figure shows that the level
crossing is reserved even when this huge disorder is introduced. 
In Fig. \ref{fig5}(c), we show that the level crossing can be destroyed
by a nonzero external potential $h$. The parameters are same as those in
(a) except $h=0.1$.

The above crossing and anti-crossing effects can be understood by
expressing the effective Hamiltonian for the low energy states (inside
the band gap) in the Majorana representation. The MFs at the joints are
denoted by $\nu_3$ and $\nu_4$ and the soliton state is regraded as a
combination of two MFs, denoted by $\nu_1$ and $\nu_2$. For the effected
Hamiltonian spanned by these $4$ Majorana states, the coupling between
$\nu_1$ and $\nu_2$ is proportional to the energy of the soliton state
$E_s$, while the coupling between $\nu_1$ and $\nu_3$ (or $\nu_2$ and
$\nu_4$) is proportional to $e^{-\alpha d_{13}}$ (or $e^{-\alpha
d_{24}}$), where $\alpha$ is proportional to the band gap of the SSH phase and
$d_{13}$ (or $d_{24}$) is the distance between the MFs $\nu_1$ and
$\nu_3$ (or $\nu_2$ and $\nu_4$). When $h=0$ and the domain wall is
moved deeply to the SSH region, the finite size effect can only lift
up the energy of the soliton state by $e^{-\alpha L}$, where $L$ is the
length of the SSH region. This is much smaller than the coupling between
$\nu_1$ and $\nu_3$ ($\nu_2$ and $\nu_4$) because $d_{13}$ and $d_{24}$
are in the scale of $\frac{L}{2}$. So the level crossing must be
protected by the leading couplings in the effective Hamiltonian.  But
when $h\ne0$, the energy of the soliton state in the SSH region is
nonzero intrinsically and the coupling between $\nu_1$ and $\nu_2$ must
be considered. So the crossing is not protected anymore. 

The above manipulation on Majorana qubit is a complementary operator to
the braiding operators. For a system with $2n$ MFs, exchanging MFs will
introduce a non-Abelian unitary transformation in the Hilbert space
spanned by the degenerated ground states. But because the braiding
operators conserve the parity of the number of fermions, they should operate in
two independent $2^{n-1}$ dimensional sub-spaces (corresponding to even
and odd parities, respectively) \cite{Ivanov2001}. By enrolling the NOT
operator we just mentioned, quantum computing can take the benefits of
the whole Hilbert space with the dimension $2^n$, instead of only half of it. 

\section{Discussions}\label{section5}

We propose a NOT operation on Majorana qubits through adiabatically
moving the complementary soliton. The two ingredients of the operation are
both topological impurities so that local disorder can not influence it.
Compared with the other proposals, our proposal is more ascendant on
flexibility and scalability. For instance, the mechanism can be extended
to a network hosting multiple Majorana qubits. One can apply the operation
on the desired qubits subsequently by moving a soliton in this network.

\acknowledgments 
The work was supported by the State Key Program for Basic Research of
China (Grant Nos. 2009CB929504, 2009CB929501), National Foundation of
Natural Science in China Grant Nos. 10704040, 11175087. 


\begin{thebibliography}{10}
\expandafter\ifx\csname url\endcsname\relax\def\url#1{\texttt{#1}}\fi

\bibitem{RevModPhys.83.1057}
\Name{Qi X.-L. \and Zhang S.-C.} \REVIEW{Rev. Mod. Phys.}{83}{2011}{1057}.

\bibitem{RevModPhys.82.3045}
\Name{Hasan M.~Z. \and Kane C.~L.} \REVIEW{Rev. Mod. Phys.}{82}{2010}{3045}.

\bibitem{RevModPhys.80.1083}
\Name{Nayak C., Simon S.~H., Stern A., Freedman M. \and Das~Sarma S.}
  \REVIEW{Rev. Mod. Phys.}{80}{2008}{1083}.

\bibitem{TIbook}
\Name{Shen S.-Q.} \Book{Topological Insulators, Dirac Equation in Condensed
  Matters} (Springer Series in Solid-State Science 174) 2012.

\bibitem{PhysRevLett.95.146802}
\Name{Kane C.~L. \and Mele E.~J.} \REVIEW{Phys. Rev. Lett.}{95}{2005}{146802}.

\bibitem{PhysRevB.75.121306}
\Name{Moore J.~E. \and Balents L.} \REVIEW{Phys. Rev. B}{75}{2007}{121306}.

\bibitem{Kitaev}
\Name{Kitaev A.~Y.} \REVIEW{Phys. Usp.}{44}{2001}{131}.

\bibitem{PhysRevLett.105.077001}
\Name{Lutchyn R.~M., Sau J.~D. \and Das~Sarma S.} \REVIEW{Phys. Rev.
  Lett.}{105}{2010}{077001}.

\bibitem{PhysRevLett.100.096407}
\Name{Fu L. \and Kane C.~L.} \REVIEW{Phys. Rev. Lett.}{100}{2008}{096407}.

\bibitem{PhysRevB.81.125318}
\Name{Alicea J.} \REVIEW{Phys. Rev. B}{81}{2010}{125318}.

\bibitem{PhysRevB.79.094504}
\Name{Sato M. \and Fujimoto S.} \REVIEW{Phys. Rev. B}{79}{2009}{094504}.

\bibitem{PhysRevLett.111.107007}
\Name{Pekker D., Hou C.-Y., Manucharyan V.~E. \and Demler E.} \REVIEW{Phys.
  Rev. Lett.}{111}{2013}{107007}.

\bibitem{PhysRevLett.106.130504}
\Name{Jiang L., Kane C.~L. \and Preskill J.} \REVIEW{Phys. Rev.
  Lett.}{106}{2011}{130504}.

\bibitem{PhysRevLett.106.130505}
\Name{Bonderson P. \and Lutchyn R.~M.} \REVIEW{Phys. Rev.
  Lett.}{106}{2011}{130505}.

\bibitem{been1}
\Name{Hassler F., Akhmerov A.~R. \and Beenakker C. W.~J.} \REVIEW{New. J.
  Phys.}{13}{2011}{095004}.

\bibitem{PhysRevLett.42.1698}
\Name{Su W.~P., Schrieffer J.~R. \and Heeger A.~J.} \REVIEW{Phys. Rev.
  Lett.}{42}{1979}{1698}.

\bibitem{PhysRevX.3.011008}
\Name{Klinovaja J. \and Loss D.} \REVIEW{Phys. Rev. X}{3}{2013}{011008}.

\bibitem{PhysRevLett.109.236801}
\Name{Klinovaja J., Stano P. \and Loss D.} \REVIEW{Phys. Rev.
  Lett.}{109}{2012}{236801}.

\bibitem{ThoulessWannier}
\Name{Thouless D.} \REVIEW{J. Phys. C}{17}{1984}{L325}.

\bibitem{PhysRevB.74.235111}
\Name{Thonhauser T. \and Vanderbilt D.} \REVIEW{Phys. Rev.
  B}{74}{2006}{235111}.

\bibitem{PhysRevLett.107.126803}
\Name{Qi X.-L.} \REVIEW{Phys. Rev. Lett.}{107}{2011}{126803}.

\bibitem{PhysRevLett.46.738}
\Name{Su W.~P. \and Schrieffer J.~R.} \REVIEW{Phys. Rev. Lett.}{46}{1981}{738}.

\bibitem{Jackiw1981253}
\Name{Jackiw R. \and Schrieffer J.} \REVIEW{Nuclear Physics B}{190}{1981}{253
  }.

\bibitem{PhysRevB.83.104522}
\Name{Santos L., Nishida Y., Chamon C. \and Mudry C.} \REVIEW{Phys. Rev.
  B}{83}{2011}{104522}.

\bibitem{explain_wallinktp}
\Book{When we use soliton state to mention the domain wall, the wall is in the
  ssh region by default.}

\bibitem{Ivanov2001}
\Name{Ivanov D.~a.} \REVIEW{Phys. Rev. Lett.}{86}{2001}{268}.

\end{thebibliography}

\end{document}